# Decarbonizing Indian Electricity Grid

Parvathy Sobha [1]

[1]Luleå Technical University, Luleå, Sweden
parvathy.sobha@ltu.se

**Abstract.** India, being one of the fastest growing economies of the world, must take a sustainable path for development. India is responsible for 7 % of global CO2 emissions. The electricity sector accounts for nearly 35% of emissions from the country. The switch from fossil fuels to renewable sources is the key in decarbonizing this sector and is considered as the crucial step for climate mitigation. This research investigates the potential of renewable energy sources (RES) - wind, solar and hydro. The optimization model developed in this study analyzes various scenarios for the transition to a sustainable future. The results show that India's aim to achieve 450 GW of installed capacity from RES is far from a Net Zero future. Results confirm that India has the potential to meet 100% of electricity demand in 2030 from RES including wind, solar and hydro. Introducing Social Cost of Carbon (SCC) is a viable option to reduce emissions in India. However, due to the low cost of coal, high coal taxes do not lead to reduced emissions.

## 1   Introduction

Currently, India is witnessing the tragic impacts of Climate change. The problem threatens the population with food insecurity, water scarcity, flooding, infectious diseases, extreme heat, economic losses, and global warming [1]-[2]. Developing countries like India face large-scale climate variability and are exposed to enhanced risks from climate change. The government of the country has conveyed that India is committed to mitigating the problem of climate change and is actively engaging in activities under UNFCC for action [3].

India, the third-largest energy consumer in the world, accounted for the emission of 2.62 billion tons of $CO_2$ in 2019 [4]. In 2020, the contribution from the electricity sector alone was 928 million tons of $CO_2$, i.e., power generation accounts for nearly 35% of emissions in India [5]. Coal power plants are majorly responsible for the high emissions in the power sector. India is the second-largest coal consumer in the world [6]. More than 50% of the power generated in the country is from coal power plants. Other fossil-based plants contribute less than 10% [7]. However, the power sector is witnessing a massive shift towards renewable energy sources (RES). The installed capacity of RES has increased by 10% during the year 2020 – 2021 and the capacity of thermal power plants were reduced by 1% (until August 2021) [8]. For each MWh of electricity produced by RES, nearly 980 kg of $CO_2$ emissions can be avoided.

The literature includes various technologies to incorporate more RES [9]. In addition, the government has introduced various policies and programs to support renewable growth in the country [10]. These policies have aided the nation to curtail $CO_2$ emissions. The government has declared that India is committed to achieve social and economic, development through sustainable ways. India has submitted Nationally Determined Contribution (NDC) to UNFCCC in 2015, which includes three main points [3]:

1. Reduce emissions by 33-35% compared to 2005 levels
2. Achieve 40% of installed capacity from RES or nuclear by 2030
3. To have a cumulative carbon sink of 2.5–3 Gigatons of $CO_2$ emissions by forest and tree cover by 2030

India also aims to include 100 GW of solar, 60 GW of wind, 10 GW of biomass, and 5 GW of hydro (small) by in generation capacity by 2022. India should meet this target while meeting increasing power demand. The country is expected to reach a population of 1.5 billion in 2030 and energy demand of 2499 TWh [3]. Being one of the fastest growing economies of the world, it is critical for India to grow sustainably.

In this paper, the author tries to identify how far India has succeeded in meeting the 2022 target, and what India should further in this decade to achieve the 2030 target. Furthermore, the possibility of including 100% RES in power generation by estimating the available potential of RES is also analyzed.

## 2 Indian Electricity Sector

Indian electricity grid is divided into five regional grids, northern, eastern, western, southern, and northeastern grid. The regional grids were operating independently before 2006. The synchronization of the regional grids was completed in 2013. Currently, India has a completely synchronized electricity grid. The advantage of the unified grid is that it opens the possibility of energy sharing between regional grids particularly with the advent of more renewable resources. Indian grid is also connected with Bhutan, Bangladesh, Myanmar, and Nepal.

### 2.1 Energy Mix

India holds the 5th largest installed capacity of RES in the world. However, 60% of installed capacity in the country accounts for fossil fuels. Table I shows the current installed capacity in India [8].

Fig. 1 shows power generation in India for the year 2020, the annual generation is around 1340 TWh. The demand is primarily met by fossil-based power plants. The generation from RES varies widely between 17 – 30% during the year. This variation is attributed to the supply from hydro plants. The contribution from hydropower increases largely during the monsoon season and decreases during summer. During the past few years, the southern states in India open shutter(s) of hydropower dams during monsoon when the water reaches Full Reservoir Level (FRL). The generation share from coal varies between 71% to 80% during the year, following variation in hydro supply. Nuclear generation remains steady around 3% – 4%.

**Table 1.** Installed capacity in India (31.08.2021)

| Fuel | Installed Capacity (GW) | % Share |
|---|---|---|
| Coal | 203 | 53 |
| Lignite | 1 | 0.2 |
| Gas | 25 | 6.5 |
| Diesel | 1 | 0.1 |
| Nuclear | 7 | 1.8 |
| Hydro | 51 | 13.4 |
| Wind | 40 | 10.4 |
| Solar | 46 | 11.9 |
| Other RES | 11 | 2.8 |

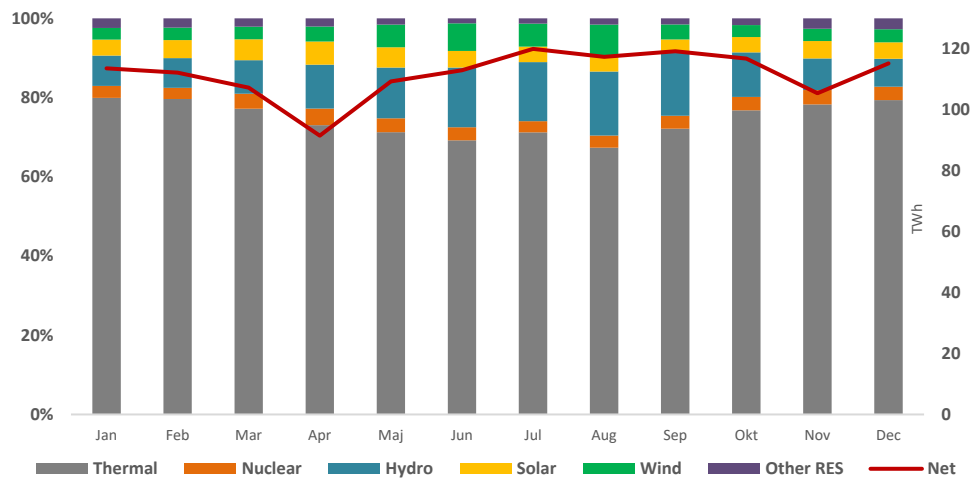

**Fig. 1.** Power Generation Mix in India, 2020

### 2.2 Energy Justice

In India, around 0.4% of the population doesn't have access to electricity (2019 data), i.e., nearly 5 million people live without electricity [11]. Moreover, the quality and reliability of electricity need to be improved in many places including both urban and rural areas. One of the 17 sustainable development goals set by UN, SDG 7 is to "ensure access to affordable, reliable, sustainable and modern energy for all". One target of SDG 7 is, "stepped-up efforts in renewable energy are needed to achieve long-term climate goals" [12]. The transition to renewable energy leads to the achievement of SDG 7.

### 2.3 Predicting Energy Supply from RES

A prediction model is employed to access the available potential of wind, solar and hydro energy for the year 2020 [13]. The average available potential of RES is assumed to remain constant over the decade. Annual generation capacity for the entire country is derived from the model at "1-hour" resolution (i.e., for 8760 hours in total). District-level data is generated from the model for better spatial resolution (state also called province is divided into different districts). The generated potential for districts is aggregated geographically to obtain renewable potential for states. Finally, the states' data are aggregated to get data of different regional grids. The determined potential for each grid is combined with the historical regional capacity factors of the grid. The final energy potential determined for each region is depicted in Fig. 2 along with a comparison to the current installed capacity. On average, only 30% of wind potential is explored in India. In the case of solar, 97 % of the potential remain unexplored.

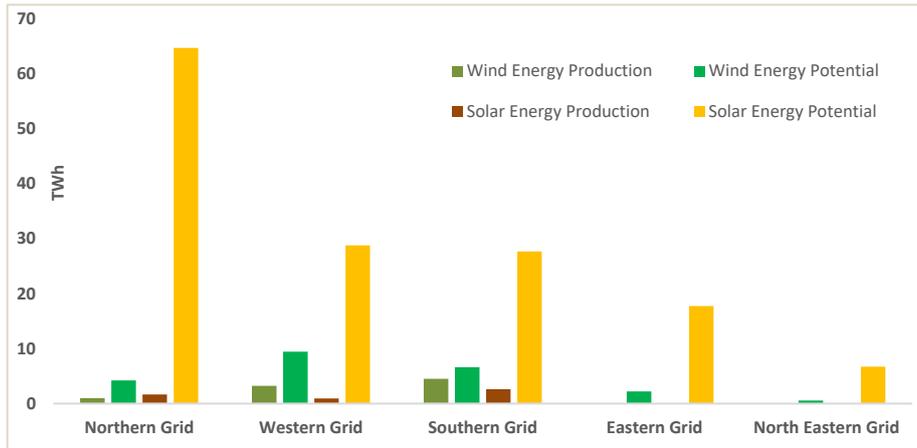

**Fig. 2.** Potential Energy Supply from RES

## 3 System Model

### 3.1 Optimization Model

In this section, a techno-economic-environmental model is developed to compare various pathways to attain reduced $CO_2$ emissions in 2030. The model is developed using single objective linear programming (LP) [14]. The model is fed with the details of existing power plants, power demand, and generation potential of RES (from the prediction model). It is fed with techno-economic & environmental related data, in-cluding emission data, fuel consumption, fuel efficiency, various plant expenditures, Levelized Cost of Energy (LCOE) [15] of different plants, etc. The purpose of the model is to determine a cost-optimized energy mix to meet the national power demand. The model can decide whether to meet demand from existing power plants or to invest in new power plants. The model uses cost, emission data, and other constraints defined to make strategic decisions regarding the electricity generation mix. The model framework is shown in Fig. 3, which describes the type of power plants, techno-economic data used, constraints employed, and the scenarios used in the model.

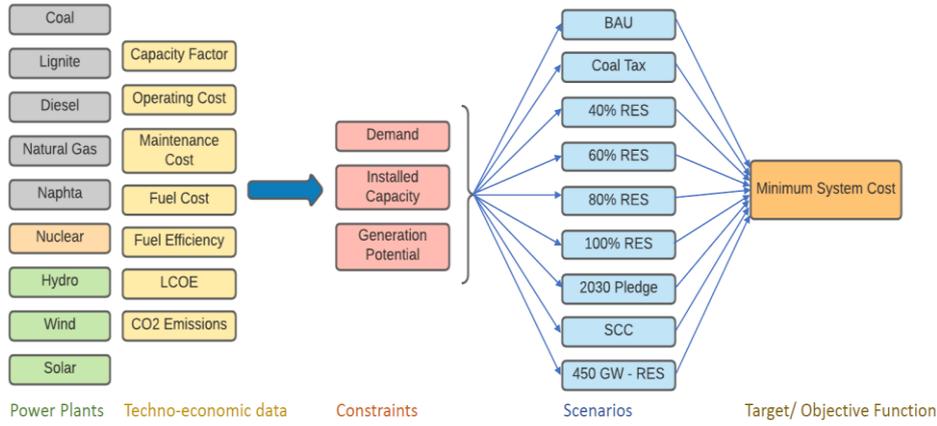

**Fig. 3** Optimization Model Framework

### 3.2    Mathematical Model

The objective function of the model is to minimize the total system cost for generating the power. In the model, similar types of power plants are aggregated for ease of computation (e.g., demand from all coal plants as summed up, generation capacity from onshore wind is aggregated). Eq. 1 describes the objective function.

$$\min \ obj.\ fn = \sum_{k=1}^{n} x_k(\alpha_k * \beta_k + \gamma_k + \delta_k + \varepsilon_k) \quad (1)$$

$x$: $generation\ from\ plant\ type\ k\ (MWh)$

$n$: $number\ of\ type\ of\ plants$

$\alpha$: $fuel\ efficiency\ (kg/MWh)$

$\beta$: $fuel\ cost\ (\$/kg)$

$\gamma$: $operating\ \&\ maintanence\ cost, fixed\ (\$/MWh)$

$\varepsilon$: $operating\ \&\ maintanence\ cost, varable\ (\$/MWh)$

$\delta$: $LCOE, calculated\ for\ new\ power\ plants\ (\$/MWh)$

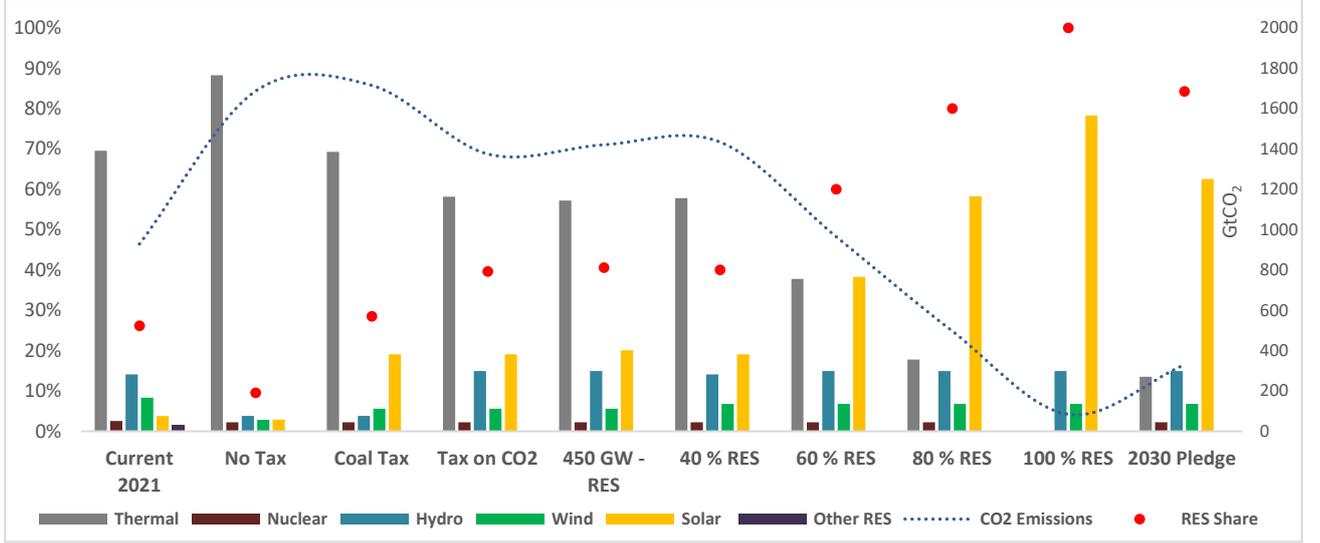

**Fig. 4** Optimization Model Framework

The constraints limiting the objective function mainly include power demand, installed capacity, and potential generation by RES. Two types of constraints are employed, one for existing plants and the other for new plants added to meet increased demand. Eq. 2, 3, and Eq. 4 describe the constraints.

$$\sum_{k=1}^{n} x_k \geq Đ \qquad (2)$$

Đ: $electricity\ demand\ (MWh)$

$$\sum_{k=1}^{n} x_k \leq Þ \qquad (3)$$

Þ: $Generation\ potential\ of\ new\ power\ plants\ (MWh)$
Þ $= f(capacity\ factor, prediction\ of\ RES, hours\ of\ operation)$

$$\sum_{k=1}^{n} x_k \leq Ć \qquad (4)$$

Ć: $Generation\ from\ existing\ power\ plants$
Ć $= f(capacity\ factor, hours\ of\ operation, installed\ capacity)$

### 3.3 Scenario Development

Scenario definition is the most significant part of the model. Scenarios explore potential futures, i.e., rather than predicting one future, scenarios explore multiple pathways to the future [16]. In the model, scenarios are defined based on both the current system and the future to which the system should evolve to. The scenarios are motivated by SDGs and India's plan to increase RES share by 2030. Table II explains the scenarios defined in the model. Here, targeted, and untargeted scenarios are considered. Targeted scenarios lead to the desired future (e.g., 40% renewables, Pledge 2030) and untargeted scenarios shows how the future will evolve for a particular situation in the present (e.g., BaU, Tax on $CO_2$).

### 3.4 Scenario Results

Scenario results are shown in Fig. 4. In the results, generation from fossil fuel-based power plants (coal, lignite, diesel, natural gas, naphtha) are aggregated as "thermal" sources. The model analyses the possibility of meeting demand with thermal, nuclear, solar, wind, and hydro energy sources. Hence the scenarios are designed likewise.

As expected, the emission curve follows RES. With high penetration of RES, emission reduces as low as 84 Gtons of $CO_2$. The share of nuclear remains constant for all the scenarios owing to the high investment cost. The share of hydro reaches a maximum of 15% and no new investments are planned. The high LCOE of hydropower restrains the model from increasing its share in the total energy mix.

TABLE I. SCENARIOS ANALYSED

| Scenario | Name | Description |
| --- | --- | --- |
| Scenario I | BAU | Business as Usual, no change in policies |
| Scenario II | Coal Tax | Tax on coal increased by 200% |
| Scenario III | 40% RES | Increase RES share to 40% of net generation |
| Scenario IV | 60% RES | Increase RES share to 60% of net generation |
| Scenario V | 80% RES | Increase RES share to 80% of net generation |
| Scenario VI | 100% RES | Increase RES share to 100% of net generation |
| Scenario VII | 2030 Pledge[a] | Decrease emission by 35% |
| Scenario VIII | SCC | Social Cost of Carbon, taxing $CO_2$ @ 20$/t$CO_2$ |
| Scenario IX | 450 GW- RES | Increase installed capacity of RES to 450 GW |

[a] 2030 Pledge is to decrease emissions from the power sector by 33-35% compared to 2005 levels

## 4 Discussions

The model determines different pathways to meet the electricity demand in 2030. Fig 4 represents the scenario results. The following section discusses the scenario results.

*Current 2021:* shows the present electricity generation mix (from January to August 2021).

*Scenario I:* The scenario analyses business as usual case, i.e., no change in current policy measures. The coal tax remains constant at 5.3 $/ton and no emission taxes are levied. No constraints are put on the energy mix. Results show that 68% of the demand is met by fossil fuels. The share of RES increased by only 4% when compared to 2021 levels.

*Scenario II:* The scenario analyses changes in the energy mix when the social cost of carbon (SCC) is included. The carbon tax was initially established in Europe and it varies from 30 cents/t$CO_2$ – 137 $/t$CO_2$ in different countries of the continent. In the analysis 20 $/t$CO_2$ is the rate chosen. In the result, fossil share reduced to 5%, and RES share increased to 93%. Levying tax on CO2 stands as a strong measure for decarbonizing the electricity sector.

*Scenario III:* Here, the tax on coal is increased by 200%. However, the fossil share is still high at 69% and renewable share is closed to 30%. Due to the low coal cost, the doubling of coal tax doesn't have any positive impact on the energy mix.

*Scenario IV:* Here, the minimum share of RES required in energy mix should be 40%. This reduced the fossil share to 58%. The policy is relevant as a short-term goal but in long term, the share of RES needs to be gradually increased.

*Scenario V:* The minimum share of RES required is 60%. The fossil share reduced to 38%. The target is relevant as a mid-term goal but in long term, the share of RES needs to be gradually increased to reduce the fossil share gradually.

*Scenario VI:* The minimum share of RES is 80% in this scenario. This reduced the fossil share to 18%. The target is highly relevant as a long-term goal to gradually phase out fossils from energy mix.

*Scenario VII:* In this scenario, the electricity demand is completely met by RES. Model results confirms that Hydro, solar, and wind energy can meet the nation's electricity demand in 2030.

*Scenario VIII:* India has pledged to reduce emissions by 33-35% compared to 2005 levels by 2030. This plan is reflected in the power sector. The scenario identifies the energy mix to reduce emissions from the power sector by 33-35%. This strategy reduces the fossil share to 14%. India will be on right track if the target is achieved by 2030.

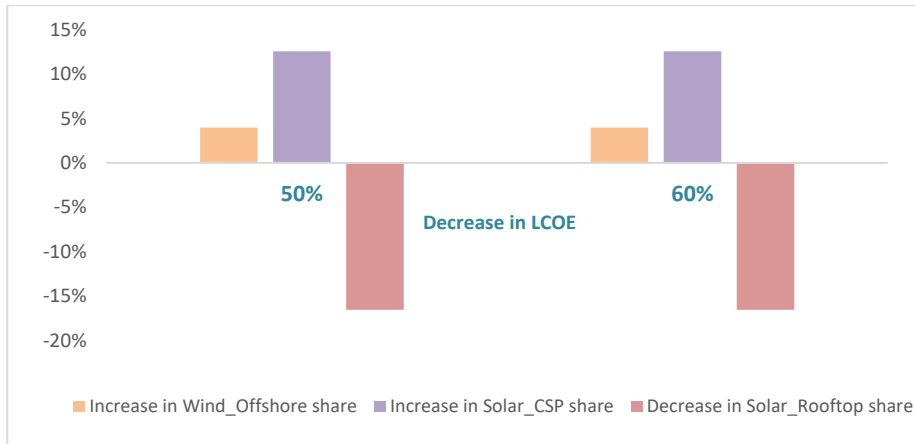

**Fig. 5** Effect of LCOE on Energy Mix

*Scenario IX:* India aims at raising the installed capacity of renewables to 450 GWs by 2030. This scenario is analyzed here. Results show that the share of fossils is 57% and that of RES is 41%. This can be a near future goal than a long-term goal, owing to high emissions from fossil fuels.

A significant finding from the analyses is the minimal or zero investment in offshore wind and concentrated solar power (CSP) plants owing to their high LCOE. Even in the scenario of 100% renewables, the share from solar PV panels (ground and rooftop mounted) reaches 78% completely excluding CSP power plants. Fig. 5 shows the increase in offshore wind and CSP share in energy mix with reduced LCOE. The results show that to increase their share, the LCOE must be reduced by at least 50%. In this case, at reduced LCOE, the maximum potential of offshore wind and CSP are utilized by the model.

Fig. 6 shows the additional capacity that needs to be installed to meet 2030 demand considering the different scenarios. The hydro capacity has to be increased to 278 GW from 46 GW. The capacity of solar PV (rooftop) must be increased between 500 – 1000 GWs. The solar panels' (ground-mounted) capacity is to be increased to 405 GWs from 38 GWs. Wind capacity (onshore) should be increased to 70 GWs from 40 GWs. The government has planned to increase RES share from 130 GWs to 450 GWs by 2030. From Fig. 10, it is evident that the target should be far higher than 450 GWs to meet power demand in 2030 with reduced emissions. With three times increase in RES in the energy mix, the share of fossils stays at 57% compared to 41% of RES. Undoubtedly this is a major achievement but the emissions from this 57% would be about 1.4 Gt of $CO_2$ for producing 1429 TWh of electricity. The annual emissions during 2019-2020 were 0,98 GtCO2 for producing 1196 TWh of electricity. In this case, the annual CO2 emissions increased by 30%. To sum up, to meet the increasing

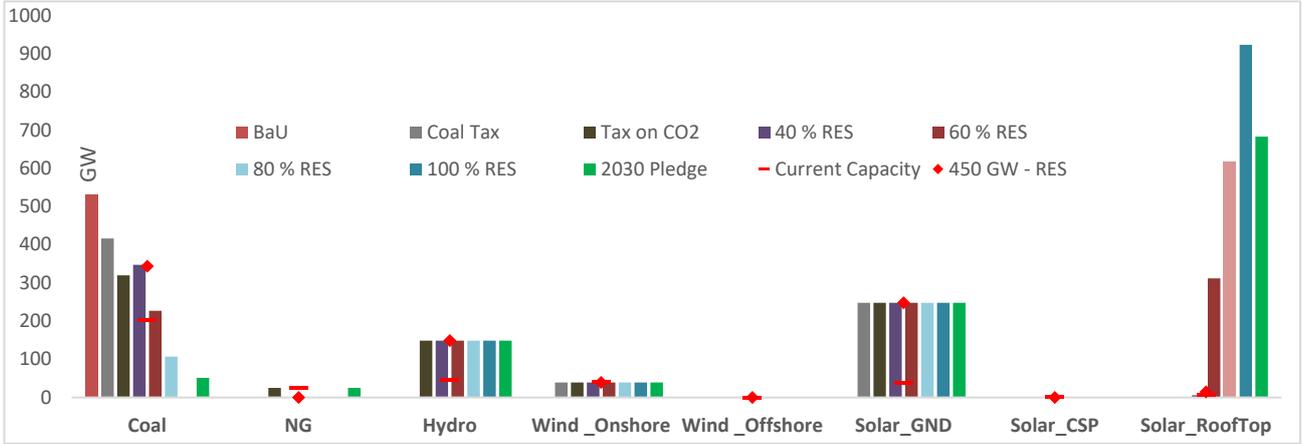

**Fig. 6** Additional Capacity for RES

energy demand with reduced emissions, India should aim at high levels of renewable integration.

## 5    Action Plan

India has been keen on promoting clean energy sources during the past decades. The renewable share has grown cumulatively by nearly 70% when compared to the beginning of the last decade. The analysis shows that there is still a large potential of RES to explore. The government has launched mitigation programs including, Green Energy Corridor for the incorporation of RES; Green Generation for Clean & Energy Secure India to achieve 175 GW of RES generation by 2022 and Solar-powered toll plazas.

Multiple projects have been rolled out to increase energy efficiency including, the National Smart Grid Mission to improve the efficiency of the power grid; Energy Conservation Campaign to reduce energy consumption by 10%; Clean Coal Policies to improve the efficiency of coal plants; National Mission for Enhanced Energy Efficiency (NMEEE) to improve energy efficiencies through policies and regulations. National Action Plan on Climate Change (NAPCC) aims at reducing per capita $CO_2$ emissions by increasing the share of RES.

Despite all the policies and programs, the renewable potential in India remains underutilized. The analysis shows that 70% of wind potential and 97% of solar potential is unused. New policies and strategies have to be designed in power domain, along with updating existing strategies. India should also aim at grid expansion policies to incorporate the growing supply to meet the increasing demand. This enables the sharing of renewable energy across different regional grids, which stands as a permanent solution for not only energy poverty and energy justice but also SDG 7.

## 6      Conclusion

India has come a long way from the last decade but has to go longer from where it stands now. Despite holding the 5th largest installed capacity of renewable energy in the world, 60% of the electricity in the country is still generated from fossil fuels. Effective strategies must be formulated along with the implementation of new technologies for reducing emissions from the electricity sector and increasing the energy efficiency of electrical equipments and appliances connected to the power grid. Above all, the country should meet the goal," Electricity for All".